# Self-organized biodiversity in biotic resource systems

Ju Kang[1], Shijie Zhang[2], Yiyuan Niu[1], Xin Wang[1*]

[1]School of Physics, Sun Yat-sen University, Guangzhou 510275, China
[2]Department of Mechanical Engineering, Massachusetts Institute of Technology, Cambridge, MA 02139, USA
[*]Correspondence: wangxin36@mail.sysu.edu.cn

## Abstract

What determines biodiversity in nature is a prominent issue in ecology, especially in biotic resource systems that are typically devoid of cross-feeding. Here, we show that by incorporating pairwise encounters among consumer individuals within the same species, a multitude of consumer species can self-organize to coexist in a well-mixed system with one or a few biotic resource species. The coexistence modes can manifest as either stable steady states or self-organized oscillations. Importantly, all coexistence states are robust to stochasticity, whether employing the stochastic simulation algorithm or individual-based modeling. Our model quantitatively illustrates species distribution patterns across a wide range of ecological communities and can be broadly used to explain biodiversity in many biotic resource systems.

## Introduction

A remarkable hallmark of life on our planet is the incredible variety of species, spanning both macroscopic and microscopic organisms that thrive in every terrestrial and aquatic niche[1-4]. Tropical forests, for instance, house a multitude of coexisting plant and vertebrate species[2], while just a gram of soil can harbor an estimated 2,000-18,000 microbial species[4]. In the sunlit zones of the oceans, approximately 150,000 eukaryotic plankton species have been identified[3]. The extraordinary diversity of coexisting species in ecosystems has posed a significant puzzle in ecology for more than half a century[1]. A major challenge arises from the well-established competitive exclusion principle (CEP), which posits that two species vying for the same type of resources cannot coexist at constant population densities[5,6], or more broadly, the number of consumer species in steady coexistence is bounded by the number of available resource species[7-9]. Interestingly, this principle appears to contradict the astonishing level of biodiversity observed in nature, particularly in scenarios such as the paradox of plankton, where a limited pool of resource types supports hundreds or more coexisting phytoplankton species[10]. So, how do plankton and many other organisms self-organize to coexist and circumvent the constraints imposed by the CEP?

Over the past several decades, various mechanisms have been proposed to challenge the limitations set by CEP and to provide explanations for biodiversity. The first category of models suggests that ecosystems never truly reach a steady state where the CEP is applicable, due to temporal variations[10,11], spatial heterogeneity[12], or self-organized dynamics[13,14]. The second category of resolutions considers factors such as cross-feeding[15-17], toxins[18], pack hunting[19], metabolic trade-offs[20,21], spatial circulation[22,23], and other interactions among the species[24-32]. In our recent study[33], we introduced a mechanistic model of pairwise encounters that extends the classical Beddington-DeAngelis (B-D) phenomenological model[25,26], and demonstrated that pairwise encounters among consumer individuals within the same species (i.e., intraspecific predator interference) promote biodiversity in abiotic resource systems.

In this work, by applying our model to biotic resource systems that typically lack cross-feeding, we show that intraspecific predator interference enables a plethora of consumer species to self-organize for coexistence with only one type of resources. Two coexistence modes are typically observed: stable steady states and self-organized oscillations. Both modes are robust to stochasticity and can thus be realized in reality. Importantly, our model quantitatively illustrates species distribution patterns across diverse ecological communities[34-36] that are highly likely to be biotic resource systems, and it can be used more broadly to explain biodiversity in many ecosystems.

## Results

**The pairwise encounter model for biotic resource systems**

We apply the mechanistic model of pairwise encounters that we introduced in a recent study[33] to biotic resource systems. Specifically, in a system where $S_C$ consumer species $\{C_1,...,C_{S_C}\}$ feed on $S_R$ biotic resource species $\{R_1,...,R_{S_R}\}$, consumer and resource individuals move around. When a consumer individual $C_i$ and a resource individual $R_l$ get close in space, the consumer can chase the resource, forming a chasing-pair $C_i^{(P)} \vee R_l^{(P)}$ (the superscript "(P)" represents pair). Meanwhile, when two consumer individuals, such as $C_i$ and $C_j$, come into proximity, they may engage in staring or fighting, thus forming an interference pair $C_i^{(P)} \vee C_j^{(P)}$ (Fig.1). In a well-mixed system where mean-field approximation is applicable, the population dynamics in the scenario involving only chasing pairs can be described as follows:

$$\begin{cases} \dot{C}_i = \sum_{l=1}^{S_R} w_{il} k_{il} x_{il} - D_i C_i, & i=1,\cdots,S_C, \\ \dot{R}_l = R_0^{(l)} R_l \left(1 - R_l/K_0^{(l)}\right) - \sum_{i=1}^{S_C} k_{il} x_{il}, & l=1,\cdots,S_R, \\ \dot{x}_{il} = a_{il} C_i^{(F)} R_l^{(F)} - (d_{il} + k_{il}) x_{il}. \end{cases} \quad (1)$$

Here, we assume that the population dynamics of the resources adheres to the same structural rule found in MacArthur's consumer-resource model[37,38]. Resources exhibit logistic growth in the absence of consumers. $R_0^{(l)}$ and $K_0^{(l)}$ denote the intrinsic growth rate and the carrying capacity of resources species $R_l$, respectively. The superscript "(F)" is used to designate populations that are freely wandering, while $x_{il} \equiv C_i^{(P)} \vee R_l^{(P)}$ represents the chasing pair. Additionally, $a_{il}$, $d_{il}$ and $k_{il}$ correspond to the encounter rate, escape rate and capture rate in the formation/dissociation of chasing pairs, respectively. Furthermore, $w_{il}$ represents the mass conversion ratio from $R_l$ to $C_i$, while $D_i$ signifies the mortality rate of $C_i$. By incorporating intraspecific predator interference, we merge Eq. 1 with the following equation:

$$\dot{y}_i = a'_i \left[C_i^{(F)}\right]^2 - d'_i y_i, \quad (2)$$

where $y_i \equiv C_i^{(P)} \vee C_i^{(P)}$ represents the intraspecific interference pair. We use $a'_{ij}$ and $d'_{ij}$ to denote the encounter rate and separate rate in the formation/dissociation of interference pairs. In this context, we set $a'_i \equiv a'_{ii}$ and $d'_i \equiv d'_{ii}$. For the case involving chasing pairs and interspecific predator interference, we integrate Eq.1 with this equation:

$$\dot{z}_{ij} = a'_{ij} C_i^{(F)} C_j^{(F)} - d'_{ij} z_{ij}, \quad i \neq j, \quad (3)$$

where $z_{ij} \equiv C_i^{(P)} \vee C_j^{(P)}$ represents the interspecific interference pair. In the scenario where chasing pairs and both intraspecific and interspecific interference are at play, we combine Eqs. 1-3. Subsequently, the population sizes of consumers and resources are represented as $C_i = C_i^{(F)} + \sum_l x_{il} + 2y_i + \sum_{j \neq i} z_{ij}$ and $R_l = R_l^{(F)} + \sum_i x_{il}$, respectively. Through the application of dimensional analysis, we non-dimensionalize all parameters as detailed in Supplemental Information (SI) Sec. 3. For ease of reference, we maintain the same notations throughout, with all parameters assumed to be dimensionless unless explicitly stated otherwise.

## Outcomes of two consumer species competing for one biotic resource species in well mixed systems

As demonstrated in our previous studies[19,33], the scenario involving only chasing pairs is subject to the constraint of CEP. To assess the role of individual mechanisms in promoting biodiversity, we conducted a comprehensive analysis of scenarios involving different types of predator interference, wherein two consumer species compete for a single biotic resource species (i.e., $S_C = 2$ and $S_R = 1$). For convenience, we choose to omit the letter "$l$" in the sub/super-script, as $S_R = 1$. Furthermore, for the sake of clarity, we define distinct competitive profiles for each consumer species, with the sole parameter displaying variation among these species being the mortality rate $D_i$. Subsequently, $\Delta \equiv (D_1 - D_2)/D_2$ represents the competitive difference between the two consumer species.

In the scenario involving chasing pairs and interspecific predator interference, our deterministic studies employing ordinary differential equations (ODEs) reveal that two consumer species cannot stably coexist at fixed points. This is due to the fact that all the fixed points for species coexistence here are inherently unstable. Meanwhile, both consumer species may persist indefinitely through time series dynamics, displaying either periodic oscillation or quasi-periodic oscillation, such as a limit cycle and a 3-D torus (see Fig. S1b-e and SI Sec. 4 for details). However, in accordance with recent findings suggesting that stochasticity can threaten species coexistence[32], the coexistence states facilitated by interspecific predator interference prove to be susceptible to stochasticity. In our study, conducted using the stochastic simulation algorithm (SSA)[39], the two consumer species fail to coexist in the presence of stochasticity (see Fig. S1).

Meanwhile, in the scenario involving chasing pairs and intraspecific interference, two types of self-organized coexistence modes are typically observed in our ODEs studies: globally attractive stable steady states, and oscillating coexistence that culminates in a limit cycle (refer to Fig. 2a-b for ODEs results, and see also Fig. S2). Notably, there is a non-zero volume of parameter space for both types of coexistence modes (see Fig. 2c-d). The fact that two consumer species can stably coexist at constant population densities with non-special parameters clearly demonstrates the violation of CEP. When considering stochasticity, remarkably, both coexistence modes facilitated by the intraspecific interference prove to be robust in the SSA studies. The two consumer species can either fluctuate around a globally attractive steady state (Fig. 2e) or exhibit sustainable stochastic oscillations in the vicinity of a stable limit cycle (Fig. 2f). It is noteworthy that the oscillating coexistence mode, which is typically vulnerable to stochasticity in other types of mechanisms[32,33], transforms into a stochastic oscillating mode of long-term coexistence within this context. Additionally, the parameter region for species coexistence in this scenario is quite similar between the SSA and ODEs studies (Fig. S3a-b).

Lastly, in the scenario involving chasing pairs and both interspecific and intraspecific interference, the species coexistence behavior closely resembles that observed without interspecific interference. Two types of previously identified coexistence modes, stable steady state and oscillating coexistence, persist, and both demonstrate robustness to stochasticity in the SSA studies (Fig. S4).

## Species coexistence modes in individual-based modeling studies

To investigate whether both long-term coexistence modes, facilitated by intraspecific interference, can be realized in practical ecosystems, we employ individual-based modeling (IBM)[40], an essentially stochastic simulation method incorporating spatial homogeneity. In our case, we consider two consumer species competing for one biotic resource species (i.e., $S_C = 2$ and $S_R = 1$) in a 2-D squared system with periodic boundary conditions of size $L^2$. In the scenario involving chasing pairs and intraspecific interference, snapshots of the system's time evolution are shown in Figs. 3a, S3c. By tracking species abundances over time, it is evident that the two consumer species can ultimately coexist with a single type of resources, exhibiting both oscillating and steady coexistence modes (Fig. 3b-e). Combining these findings with the SSA simulations, it becomes evident that intraspecific predator interference can promote species coexistence in both long-term coexistence modes, irrespective of stochasticity.

**Self-organized biodiversity in systems containing multiple consumer species**

We proceed to investigate the generic case in which multiple consumer species compete for a limited biotic resource species ($S_C > S_R$) in a scenario involving chasing pairs and intraspecific interference. Thus, the population dynamics is governed by Eqs. 1-2. Similar to the cases above, we assign each consumer species of unique competitiveness through a distinctive $D_i$ $(i = 1, \cdots, S_C)$. Remarkably, a diverse array of consumer species may self-organize to coexist with only one or a few types of biotic resources in both oscillating and steady coexistence modes (Figs. 4 and S5). Furthermore, both coexistence modes are robust to stochasticity, as we confirm the results using SSA simulations (Figs. 4 and S5). In fact, both coexistence modes can be switched by tuning the separate rate $d_i'$ in the interference pair. As the separate rate $d_i'$ increases, the self-organized biodiversity mode changes from stable steady coexistence into oscillating coexistence. Similar results are observed in the case of $S_C = 2$ and $S_R = 1$ (Fig. S2a-b). Combining these analyses (see Figs. 2, 4, and S2), we identify a deterministic/stochastic Hopf bifurcation in the ODEs/SSA studies, respectively.

When the population size of each resource species greatly exceeds the total population of consumers (i.e., $R_l \gg \sum_{i=1}^{S_C} C_i$, with $l = 1, \cdots, S_R$), it becomes possible to determine the steady-state populations of both consumer and resource species analytically (see SI Sec. 2 for details). As shown in Figs. S2c, S5a-b, the analytical solutions are highly consistent with the numerical results.

**Species distributions in ecological communities**

The quantitative aspect of biodiversity in an ecological community is manifested through the species distribution pattern. Most notably, the rank-abundance curves of different ecological communities exhibit a universally similar shape. We collect ecological data from a wide range of communities that potentially feed on biotic resource species found in existing literature[34-36], including birds, bats, fish, and lizards. Next, we apply our model of biotic resource systems in the scenario involving chasing pairs and intraspecific interference to simulate ecological communities in a well-mixed case, where a single resource species ($S_R = 1$) supports a plethora of consumer species ($S_C \gg S_R$). In our model settings for each ecological community, consumer species are assigned unique competitiveness levels as aforementioned, while the mortality rates of consumers follow a Gaussian distribution with a coefficient of variance (CV) taken around 0.3[41] (see SI Sec. 5.2 for details).

The comparison of species distributions between numerical results and experimental data in diverse ecological communities is presented in Fig. 5. Regarding the rank-abundance curves (Fig. 5a), both the deterministic results obtained through ODEs and stochastic results derived from SSA agree quantitatively with experimental observations. Simulations results and experimental data exhibit similar Shannon entropies (see SI Sec. 5.2), and are largely considered statistically identical distributions according to the Kolmogorov-Smirnov (K-S) test (using a significance threshold of 0.05). Nonetheless, a discernible disparity emerges for species with low relative abundance within certain communities (Fig. 5b), which could potentially be attributed to population drift within real communities.

## Discussions

The inconsistency between CEP and biodiversity has been a significant challenge in ecology for many years. Despite many mechanisms aimed at circumventing the constraints imposed by CEP[5,7,8,10-15,18-22,24-32], it remains elusive how ecosystems maintain species diversity in well-mixed biotic resource systems that typically lack cross-feeding. By applying our established mechanistic model[33] of pairwise encounters in biotic resource systems, and using a combination of numerical simulations and mathematical analysis, we reveal that the intraspecific interference among consumers enables a multitude of consumer species to self-organize for coexistence in two modes: oscillating coexistence and stable steady coexistence. Both coexistence modes are robust to stochasticity and can, therefore, be realized in practical ecosystems. Applying the above analysis to real ecological systems, our model quantitatively illustrates species distributions for a broad range of ecological communities.

Predator inference was originally introduced in the B-D model long ago[25,26]. Nevertheless, the functional response within the B-D model that encompasses intraspecific interference can be formally deduced from the scenario involving only chasing pairs without predator interference[19,42]. Consequently, this raises substantial concern when applying the B-D model to challenge CEP. Therefore, we employ our established mechanistic model[33] for biotic resource systems in this study.

Lastly, it is known that stochasticity tends to jeopardize species coexistence, especially in cases of deterministic oscillating coexistence[32,33]. For pairwise encounters between individuals in biotic resource systems, the oscillating coexistence promoted by interspecific interference is disrupted by stochasticity. In contrast, intraspecific predator interference can facilitate both stochastic oscillating coexistence around a limit cycle and stochastic semi-steady coexistence around a fixed point. This has the potential to be broadly used to illustrate species coexistence patterns and dynamics in natural ecosystems.

## Acknowledgements

We thank Roy Kishony, Eric D. Kelsic, Yang-Yu Liu and Fan Zhong for helpful discussions. This work was supported by National Natural Science Foundation of China (Grant No.12004443), Guangzhou Municipal Innovation Fund (Grant No.202102020284) and the Hundred Talents Program of Sun Yat-sen University.

## Author contributions

X.W. conceived and designed the project; J.K., S.Z., Y.N. and X. W. performed research; J.K. and X. W. analyzed data and wrote the paper.

## Competing interests

The author declares no competing interests.

## Data and code availability

All study data are included in the article and/or Supplemental Information.

**Figures**

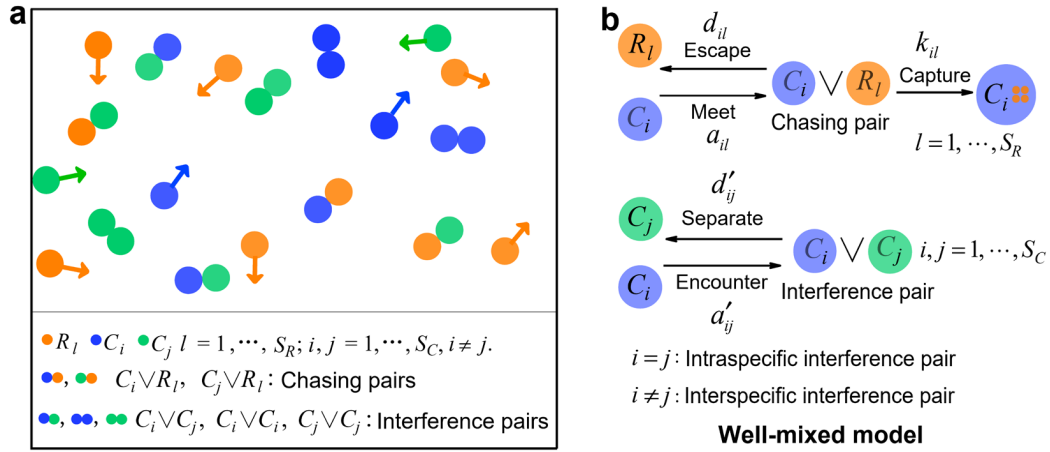

**Fig. 1 | A model of pairwise encounter for biotic resource systems.** (**a**) A schematic of $S_C$ consumer species competing for $S_R$ resources species. (**b**) The model from (**a**) in a well-mixed system.

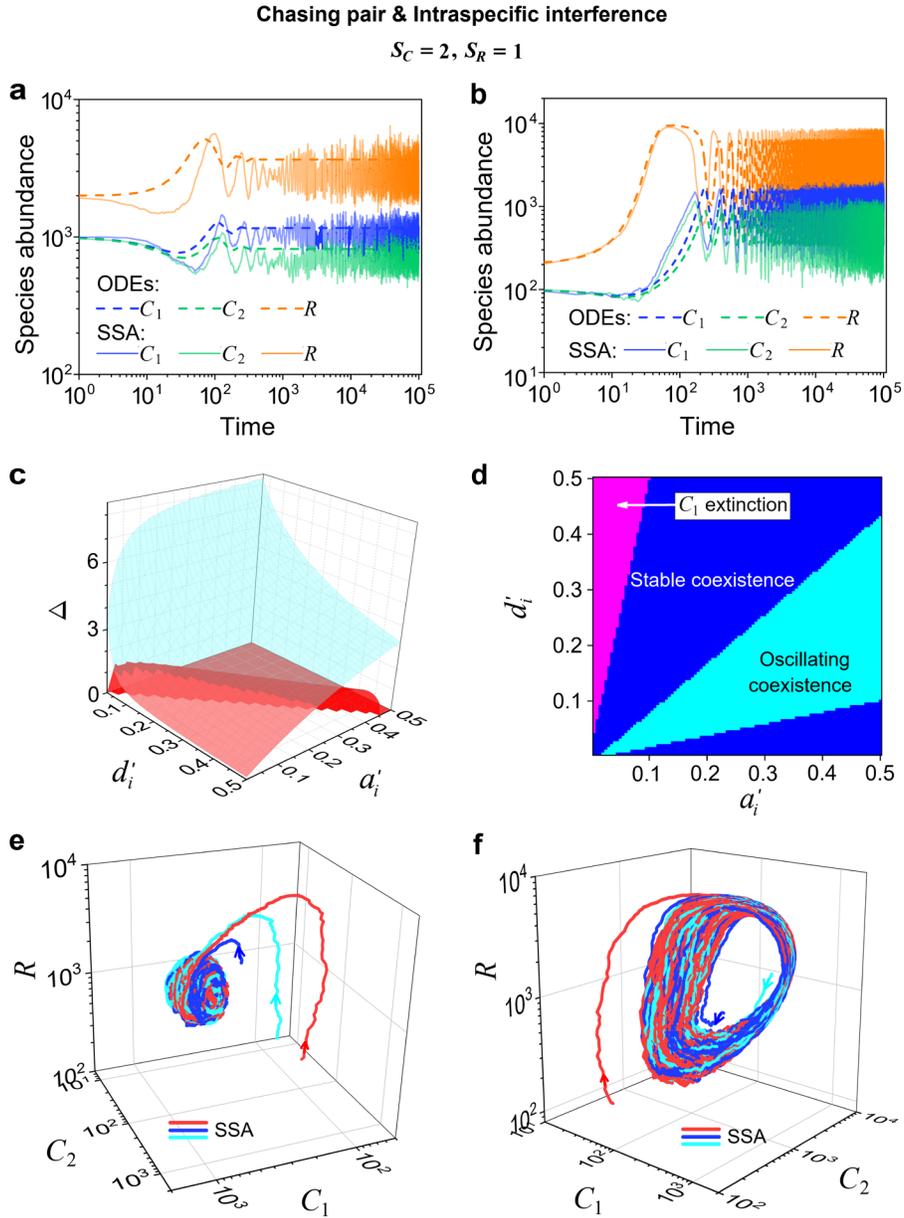

**Fig. 2 | Long-term coexistence modes of two consumer species with one biotic resource species.** Here $D_i$ ($i=1,2$) is the only parameter that varies with the two consumer species, and $\Delta \equiv (D_1 - D_2)/D_2$ measures the competitive difference between the two species. (**a**, **e**) Stable coexistence. (**b**, **f**) Coexistence in an oscillating mode. (**a-b**) Representative time courses of species abundances simulated using ordinary differential equations (ODEs) or stochastic simulation algorithm (SSA). (**e-f**) Representative trajectories of species coexistence in the phase space simulated using SSA (refer to (**a-b**) for the time courses). (c) A 3D phase diagram in the ODEs study. The region below the blue surface and above the red surface represents stable coexistence. The region below the red surface and above $\Delta = 0$ represents unstable fixed points. (d) The transection that corresponds to the $\Delta = 1$ plane in (c).

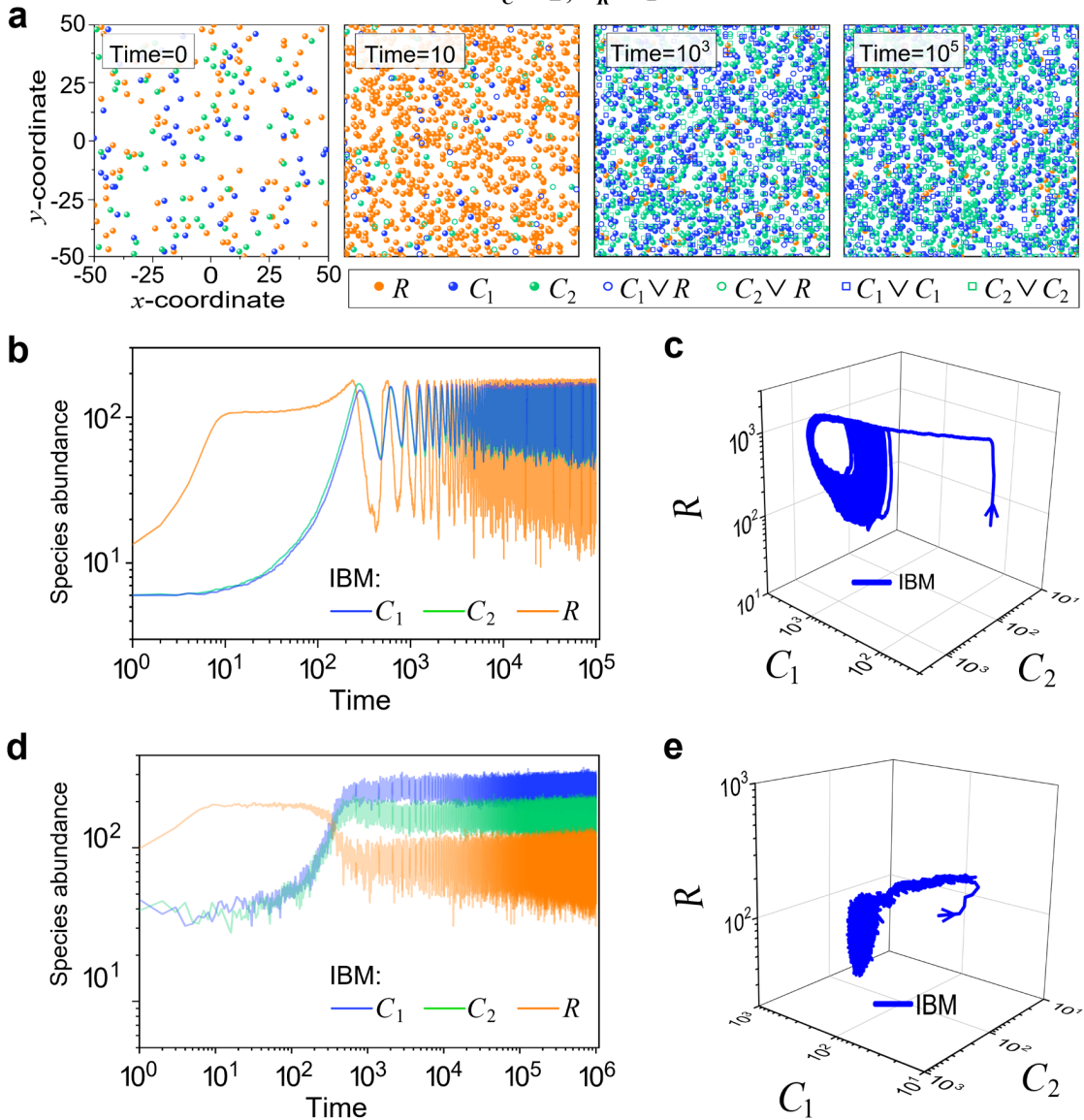

**Fig. 3 | Stable and oscillating coexistence spatial patterns of two consumer species competing for a single biotic resource species.** (**a**) Snapshots of the individual-based modeling (IBM). (**b, d**) Representative time courses of species abundances simulated with IBM. (**c, e**) Representative trajectories in the phase space ending in a stochastic limit cycle or stable state (refer to (**b**) and (**d**) for the time courses). See Fig. S3c for snapshots of the stable coexistence mode using the IBM simulation.

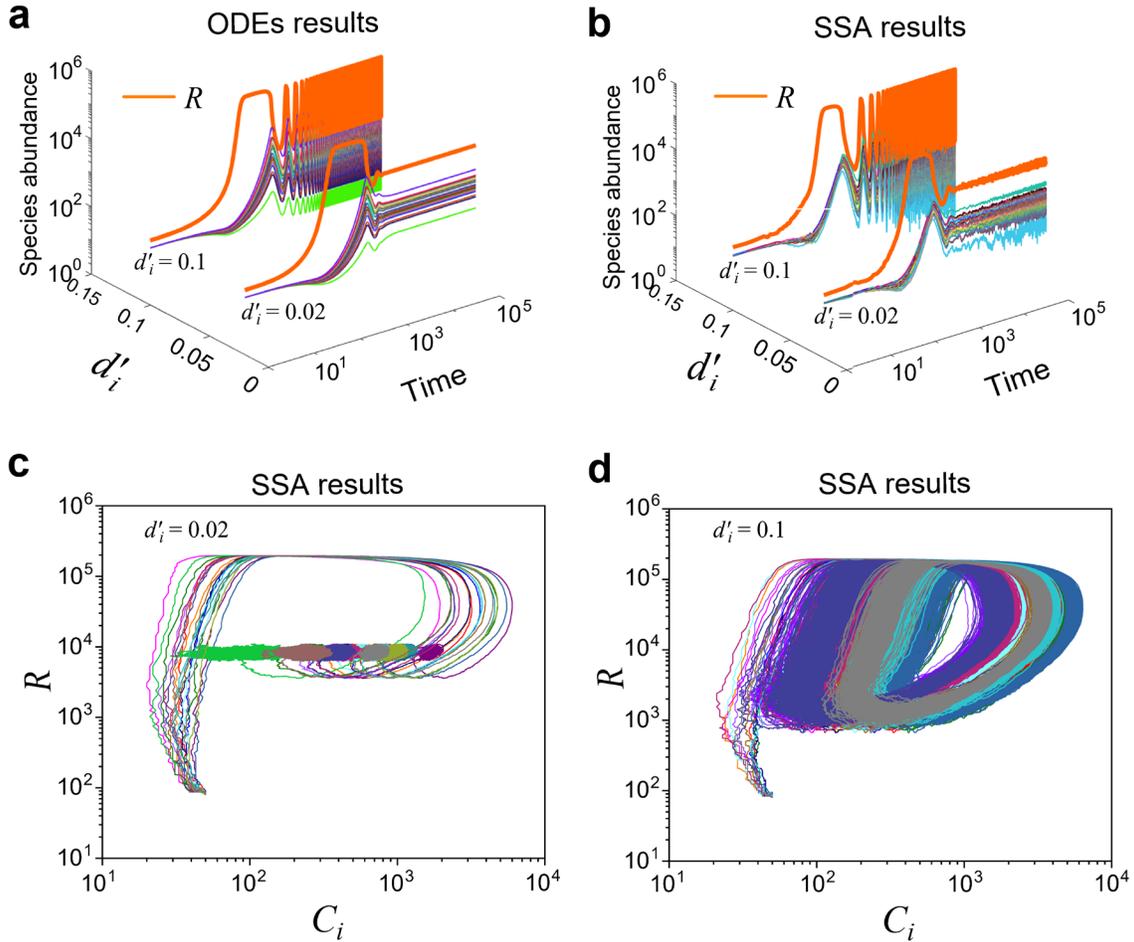

**Fig. 4 | The separate rate within the interference pair alters self-organized biodiversity modes.** Here, $D_i$ is the only parameter that varies with the consumer species, which was randomly drawn from a Gaussian distribution. (**a-b**) Time courses of species abundances simulated with ODEs or SSA. (**c-d**) Representative trajectories of species coexistence in 2-D projections of the phase space, simulated with SSA (refer to (**a-b**) for the time courses).

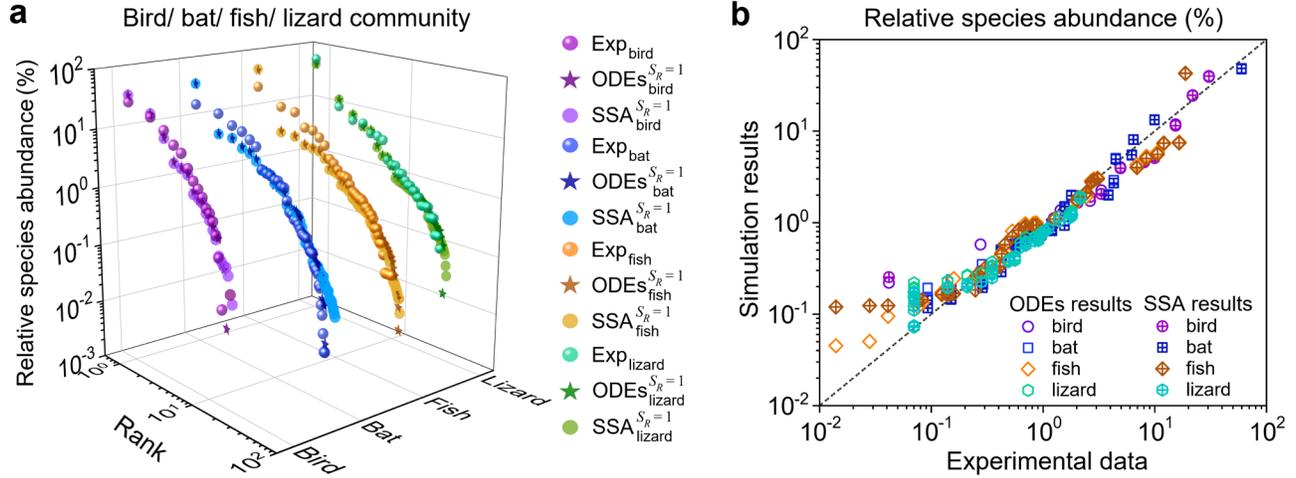

**Fig. 5 | Explanation of the species distribution patterns across diverse ecological communities.** (**a**) The rank-abundance patterns in four different communities. The 3-D spherical icons represent the experimental data (marked with "Exp") obtained from previous studies[34-36]. The ODEs and SSA results were derived from a timestamp at $t = 10^5$ within the time series (see Fig. S5). In the model settings for each ecological community, multiple consumer species coexist with only one type of biotic resource species ($S_R = 1$), with $S_C$ values of 20 (bird), 40 (bat), 50 (fish), and 55 (lizard), respectively. $D_i$ $(i = 1, \cdots, S_C)$ is the only parameter that varies with the consumer species and was randomly drawn from a Gaussian distribution. In the Kolmogorov-Smirnov (K-S) test, the probabilities (*p*-values) indicating whether the simulation results and the corresponding experimental data come from identical distributions are as follows: $p_{\text{ODEs}}^{\text{bird}} = 0.81$, $p_{\text{SSA}}^{\text{bird}} = 0.36$; $p_{\text{ODEs}}^{\text{bat}} = 0.94$, $p_{\text{SSA}}^{\text{bat}} = 0.28$; $p_{\text{ODEs}}^{\text{fish}} = 0.98$, $p_{\text{SSA}}^{\text{fish}} = 0.18$; $p_{\text{ODEs}}^{\text{lizard}} = 0.42$, $p_{\text{SSA}}^{\text{lizard}} = 0.32$. Using a significance threshold of 0.05, none of the *p*-values suggest the existence of a statistically significant difference. (**b**) An intuitive comparison of species distribution in the four communities, as depicted in (**a**), between model results and experimental data.

**Supplemental Information**

# Self-organized biodiversity in biotic resource systems

Ju Kang[1], Shijie Zhang[2], Yiyuan Niu[1], Xin Wang[1*]

1. *School of Physics, Sun Yat-sen University, Guangzhou* 510275, *China.*
2. *Department of Mechanical Engineering, Massachusetts Institute of Technology, Cambridge, MA 02139, USA*
[*]Correspondence: wangxin36@mail.sysu.edu.cn

**Contents**

# 1. Steady-state analytical solutions for two consumer species competing for one biotic resource species

We consider the scenario involving chasing pairs and intraspecific interference in the simple case of $S_C = 2$ and $S_R = 1$: The populations of consumers and resources are $C_i = C_i^{(F)} + x_i + 2y_i$ and $R = R^{(F)} + \sum_{i=1}^{2} x_i$, respectively. Therefore, the population dynamics of the consumers and resources can be described as follows:

$$\begin{cases} \dot{x}_i = a_i R^{(F)} C_i^{(F)} - (d_i + k_i)x_i, \quad i=1,2. \\ \dot{y}_i = a_i' \left[ C_i^{(F)} \right]^2 - d_i' y_i, \\ \dot{C}_i = w_i k_i x_i - D_i C_i, \\ \dot{R} = R_0 R (1 - R/K_0) - \sum_{i=1}^{2} k_i x_i. \end{cases} \quad (S1.1)$$

For simplicity, we define $K_i \equiv (d_i + k_i)/a_i$, $\alpha_i \equiv D_i/(w_i k_i)$, and $\beta_i \equiv a_i'/d_i'$ $(i=1,2)$.

At steady state, since $\dot{x}_i = 0$, $\dot{y}_i = 0$ and $\dot{C}_i = 0$ ($i=1,2$), we have

$$\begin{cases} x_i = \alpha_i C_i, \\ C_i^{(F)} = K_i \alpha_i C_i / R^{(F)}, \\ y_i = \beta_i (K_i \alpha_i C_i)^2 \left[ R^{(F)} \right]^{-2}, \end{cases} \quad (S1.2)$$

Meanwhile, $C_i = C_i^{(F)} + x_i + 2y_i$ and $C_i, R > 0$ ($i=1,2$), thus:

$$C_i = \frac{(1-\alpha_i)\left[R^{(F)}\right]^2 - K_i \alpha_i R^{(F)}}{2\beta_i (K_i \alpha_i)^2}. \quad (S1.3)$$

If the total population size of the resources is much larger than that of the consumers (i.e., $R \gg C_1 + C_2$), then $R \gg x_1 + x_2$ and $R^{(F)} \approx R$. Consequently,

$$C_i \approx \frac{(1-\alpha_i) R^2 - K_i \alpha_i R}{2\beta_i (K_i \alpha_i)^2}. \quad (S1.4)$$

With $\dot{R} = 0$, then,

$$R = \frac{k_1/(2\beta_1 K_1) + k_2/(2\beta_2 K_2) + R_0}{\dfrac{k_1(1-\alpha_1)}{2\beta_1 \alpha_1 (K_1)^2} + \dfrac{k_2(1-\alpha_2)}{2\beta_2 \alpha_2 (K_2)^2} + \dfrac{R_0}{K_0}}. \quad (S1.5)$$

Eqs. S1.4-S1.5 are the analytical solutions to the steady-state species abundances for the case of $R \gg C_1 + C_2$. As shown in Fig. S2c, the analytical solutions agree well with the numerical results (the exact solutions).

Next, we use linear stability analysis to analyze the local stability of the fixed point. Specifically, for an arbitrary fixed point $E(x_1, x_2, y_1, y_2, C_1, C_2, R)$, only when all the eigenvalues (defined as $\lambda_i = 1, \cdots, 7$) of the Jacobian matrix at point $E$ have negative real parts, would the point be locally stable. In this study, we observe that as the separation rate $d'$ $(d' = d'_1 = d'_2)$ of the interference pair increases near the critical value $d'_c$, the population dynamics of the system transits from a stable fixed point into a stable limit cycle (Fig. 2a-b, see also Fig. S2). Notably, the amplitude of the oscillation is directly proportional to $\sqrt{d' - d'_c}$ (Fig. S2a-b), providing clear evidence of a supercritical Hopf bifurcation[1].

## 2. Steady state analytical solutions for $S_C$ consumer species competing for $S_R$ biotic resource species

Here we consider the scenario involving chasing pair and intraspecific interference for the generic case with $S_C$ types of consumers and $S_R$ types of biotic resources. Then, the population dynamics of the consumers and resources can be described as follows:

$$\begin{cases} \dot{x}_{il} = a_{il} C_i^{(F)} R_l^{(F)} - (d_{il} + k_{il}) x_{il}, \\ \dot{y}_i = a'_{ii} \left[ C_i^{(F)} \right]^2 - d'_{ii} y_i, \\ \dot{C}_i = \sum_{l=1}^{S_R} w_{il} k_{il} x_{il} - D_i C_i, \\ \dot{R}_l = R_0^{(l)} R_l \left(1 - R_l / K_0^{(l)}\right) - \sum_{i=1}^{S_C} k_{il} x_{il}, \quad i = 1, \cdots, S_C, \; l = 1, \cdots, S_R. \end{cases} \quad \text{(S2.1)}$$

Note that Eq. S2.1 is identical to Eq. 1-2, and we use the same variables and parameters as that in the main text. Then, the populations of the consumers and resources are $C_i = C_i^{(F)} + \sum_{l=1}^{S_R} x_{il} + 2 y_i$ and $R_l = R_l^{(F)} + \sum_{i=1}^{S_C} x_{il}$, respectively. For convenience, we define $K_{il} \equiv (d_{il} + k_{il}) / a_{il}$, $\alpha_{il} \equiv D_i / (w_{il} k_{il})$, and $\beta_i \equiv a'_{ii} / d'_{ii}$ $(i = 1, \cdots, S_C, \; l = 1, \cdots, S_R)$.

At steady state, from $\dot{x}_{il} = 0$, $\dot{y}_i = 0$, and $\dot{C}_i = 0$, we have

$$\begin{cases} x_{il} = C_i^{(F)} R_l^{(F)} / K_{il}, \\ y_i = \beta_i \left[ C_i^{(F)} \right]^2, \\ C_i = \sum_{l=1}^{S_R} x_{il} / \alpha_{il} = \sum_{l=1}^{S_R} C_i^{(F)} R_l^{(F)} / (K_{il} \alpha_{il}). \end{cases} \quad \text{(S2.2)}$$

Meanwhile, $C_i = C_i^{(F)} + \sum_{l=1}^{S_R} x_{il} + 2y_i$. Note that $C_i^{(F)} > 0$, thus:

$$C_i^{(F)} = \frac{1}{2\beta_i}\left[-1 + \sum_{l=1}^{S_R}\left(\frac{1}{\alpha_{il}} - 1\right)\frac{R_l^{(F)}}{K_{il}}\right]. \tag{S2.3}$$

Combined with Eq. S2.3, and then

$$C_i = \sum_{l=1}^{S_R} \frac{R_l^{(F)}}{2\beta_i \alpha_{il} K_{il}}\left[-1 + \sum_{l'=1}^{S_R}\left(\frac{1}{\alpha_{il'}} - 1\right)\frac{R_{l'}^{(F)}}{K_{il'}}\right]. \tag{S2.4}$$

Combining with $\dot{R}_l = 0$, we have

$$R_0^{(l)} R_l \left(1 - \frac{R_l}{K_0^{(l)}}\right) = \sum_{i=1}^{S_C} \frac{k_i}{2\beta_i K_{il}}\left[-1 + \sum_{l'=1}^{S_R}\left(\frac{1}{\alpha_{il'}} - 1\right)\frac{R_{l'}^{(F)}}{K_{il'}}\right] R_l^{(F)}. \tag{S2.5}$$

If the population size of each resource species is much larger than the total population size of all consumers (i.e., $R_l \gg \sum_{i=1}^{S_C} C_i$, $l = 1, \cdots, S_R$), then $R_l \gg \sum_{i=1}^{S_C} x_{il}$, and $R_l^{(F)} \approx R_l$.

Since $R_l > 0$ $(l = 1, \cdots, S_R)$, then

$$\sum_{l'=1}^{S_R}\left[\delta_{l,l'}\frac{R_0^{(l)}}{K_0^{(l)}} + \sum_{i=1}^{S_C}\frac{k_{il}}{2\beta_i K_{il} K_{il'}}\left(\frac{1}{\alpha_{il'}} - 1\right)\right]R_{l'} = R_0^{(l)} + \sum_{i=1}^{S_C}\frac{k_{il}}{2\beta_i K_{il}}, \tag{S2.6}$$

where $\delta_{l,l'} = \begin{cases} 0, & l \neq l'. \\ 1, & l = l'. \end{cases}$ To present Eq. S2.6 in a matrix form, we define matrix

$\mathbf{A} \equiv [A_{sq}] \in \mathbb{R}^{S_R \times S_R}$ ($\mathbb{R}$ stands for the real number field), with

$$A_{sq} = \delta_{s,q}\frac{R_0^{(s)}}{K_0^{(s)}} + \sum_{i=1}^{S_C}\frac{k_{is}}{2\beta_i K_{is} K_{iq}}\left(\frac{1}{\alpha_{iq}} - 1\right), \quad s, q = 1, \cdots, S_R, \tag{S2.7a}$$

and two arrays

$$\begin{cases} \mathbf{B} \equiv (R_0^{(1)} + \sum_{i=1}^{S_C}\frac{k_{i1}}{2\beta_i K_{i1}}, \cdots, R_0^{(S_R)} + \sum_{i=1}^{S_C}\frac{k_{iS_R}}{2\beta_i K_{iS_R}})^T, \\ \mathbf{R} \equiv (R_1, \cdots, R_{S_R})^T, \end{cases} \tag{S2.7b}$$

where "T" represents the transpose. Then, Eq. S2.6 can be written as

$$\mathbf{A} \cdot \mathbf{R} = \mathbf{B}. \tag{S2.7c}$$

Then, we can solve for $\mathbf{R}$, with

$$\mathbf{R} = \frac{\text{adj}(\mathbf{A})}{\det(\mathbf{A})} \cdot \mathbf{B}, \tag{S2.7d}$$

where $\text{adj}(\mathbf{A})$ and $\det(\mathbf{A})$ denote the adjugate matrix and determinant of $\mathbf{A}$,

respectively. Next, we define $\mathbf{C} \equiv (C_1, \cdots, C_{S_C})$. Note that $R_l \approx R_l^{(F)}$, combining with Eq. S2.4, we have

$$\begin{cases} \mathbf{C} \equiv (C_1, \cdots, C_{S_C}), \\ C_i = \sum_{l=1}^{S_R} \frac{R_l}{2\beta_i \alpha_{il} K_{il}} \left[ -1 + \sum_{l'=1}^{S_R} \left( \frac{1}{\alpha_{il'}} - 1 \right) \frac{R_{l'}}{K_{il'}} \right], \quad i = 1, \cdots, S_C. \end{cases} \quad (S2.7e)$$

Eqs. S2.7a-S2.7e are the analytical solutions to the steady-state species abundances for the case of $R_l \gg \sum_{i=1}^{S_C} C_i$ $(l = 1, \cdots, S_R)$. For the case of $S_R = 1$, we can explicitly present the analytical solution of the steady-state species abundances. To simplify the notations, we omit the "$l$" in the sub-/super-scripts since $S_R = 1$. Then,

$$\begin{cases} R = \dfrac{R_0 + \sum_{i=1}^{S_C} k_i / (2\beta_i K_i)}{\sum_{i=1}^{S_C} \dfrac{k_i (1-\alpha_i)}{2\beta_i \alpha_i (K_i)^2} + \dfrac{R_0}{K_0}}, \\ C_i = \dfrac{1}{2\beta_i \alpha_i K_i} \left[ \left( \dfrac{1}{\alpha_i} - 1 \right) \dfrac{R}{K_i} - 1 \right] R, \quad i = 1, \cdots, S_C. \end{cases} \quad (S2.8)$$

Numerically, consistent with the analytical predictions, a handful of resource species can sustain a remarkably diverse array of coexisting consumer species ($S_C \gg S_R$) in a stable state (Figs. S5). The comparisons between the analytical predictions and the ODE simulation results (exact solution) are shown in Figs. S2c and S5a-b, demonstrating clear and strong consistency.

3. **Dimensional analysis of biotic resource systems**

The population dynamics of the scenario, encompassing chasing pairs and both intra- and inter-specific interference, can be described as follows:

$$\begin{cases} \dot{x}_{il} = a_{il} C_i^{(F)} R_l^{(F)} - (d_{il} + k_{il}) x_{il}, \\ \dot{y}_i = a'_{ii} \left[ C_i^{(F)} \right]^2 - d'_{ii} y_i, \\ \dot{z}_{ij} = a'_{ij} C_i^{(F)} C_j^{(F)} - d'_{ij} z_{ij}, \\ \dot{C}_i = \sum_{l=1}^{S_R} w_{il} k_{il} x_{il} - D_i C_i, \\ \dot{R}_l = R_0^{(l)} R_l \left( 1 - R_l / K_0^{(l)} \right) - \sum_{i=1}^{S_C} k_{il} x_{il}, \end{cases} \quad (S3.1)$$

with $i, j = 1, \cdots, S_C$, $l = 1, \cdots, S_R$, and $i \neq j$. Here, $C_i$ and $R_l$ represent the population sizes of consumers and resources in the system, with $C_i = C_i^{(F)} + \sum_l x_{il} + 2y_i + \sum_{j \neq i} z_{ij}$ and $R_l = R_l^{(F)} + \sum_i x_{il}$. Notably, Eq S3.1 already incorporates dimensionless variables and parameters: $x_{il}$, $y_i$, $z_{ij}$, $C_i^{(F)}$, $R_l^{(F)}$, $C_i$, $R_l$, $w_{il}$, and $K_0^{(l)}$. To nondimensionalize all the terms, we define $\tilde{t} = t / \tau$, where $\tau = \tilde{D}_1 / D_1$ and $\tilde{D}_1$ is a reducible dimensionless parameter that can be freely chosen with any positive value. Next, we introduce dimensionless parameters $\tilde{a}_{il} = a_{il}\tau$, $\tilde{d}_{il} = d_{il}\tau$, $\tilde{k}_{il} = k_{il}\tau$, $\tilde{a}'_i = a'_i\tau$, $\tilde{d}'_i = d'_i\tau$, $\tilde{a}'_{ij} = a'_{ij}\tau$, $\tilde{d}'_{ij} = d'_{ij}\tau$, $\tilde{D}_i = \tau D_i$ and $\tilde{R}_0^{(l)} = \tau R_0^{(l)}$. Substituting these dimensionless terms into Eq. S3.1 yields the population dynamics in dimensionless form:

$$\begin{cases} \dot{x}_{il} = \tilde{a}_{il} C_i^{(F)} R_l^{(F)} - \left(\tilde{d}_{il} + \tilde{k}_{il}\right) x_{il}, \\ \dot{y}_i = \tilde{a}'_{ii} \left[C_i^{(F)}\right]^2 - \tilde{d}'_{ii} y_i, \\ \dot{z}_{ij} = \tilde{a}'_{ij} C_i^{(F)} C_j^{(F)} - \tilde{d}'_{ij} z_{ij}, \\ \dot{C}_i = \sum_{l=1}^{S_R} w_{il} \tilde{k}_{il} x_{il} - \tilde{D}_i C_i, \\ \dot{R}_l = \tilde{R}_0^{(l)} R_l \left(1 - R_l / K_0^{(l)}\right) - \sum_{i=1}^{S_C} \tilde{k}_{il} x_{il}. \end{cases} \qquad (S3.2)$$

For simplicity, we omit the notation "~" and use dimensionless variables and parameters in simulation studies, unless explicitly mentioned otherwise.

### 4. Identification of quasi-periodic oscillation through numerical analysis

To identify the type of nonlinear dynamics depicted in Fig. S1c and e, we employ the Poincaré map and the Lyapunov exponents in our analysis (Fig. S1f-g). For the case of two consumer species competing for one biotic resource species involving chasing pairs and interspecific interference, the population dynamics can be described by Eqs. 1 and 3 with $S_C = 2$ and $S_R = 1$, where there are 6 independent variables, namely, $x_1$, $x_2$, $z$, $C_1$, $C_2$, $R$. Qualitatively, the Poincaré map presents closed curves consisting of a finite number of points (Fig. S1g), indicating quasi-periodic motion[2]. Quantitatively, among the 6 Lyapunov exponents following the definition by Wolf et al.[3], as shown in Fig. S1f, three of the Lyapunov exponents are zero, while the rest are all negative, clearly demonstrating a 3-D torus[2].

### 5. Numerical simulation settings

### 5.1 Individual-based modeling

To mimic real ecosystems, we employ individual-based modeling (IBM)[4,5]. In the IBM studies, we examine a 2D square system with dimensions of $L^2$ and apply periodic boundary conditions. In a system where two consumer species competing for one biotic resource species, consumer individuals of species $C_i$ $(i=1,2)$ move at a speed $v_{C_i}$, while the resources move at a speed $v_R$. Here, the unit length is set to $\Delta l = 1$, and all individuals move stochastically. More precisely, when $\Delta t$ is sufficiently small, such that $v_{C_i} \Delta t \ll 1$. $C_i$ individuals jump a unit length with the probability $v_{C_i} \Delta t$. In practice, we simulate the temporal evolution of the model system using the following procedures.

(1) *Initialization*: Initial positions are randomly selected from a uniform distribution in the squared space and rounded to the nearest integer point in the $x-y$ coordinates.

(2) *Moving*: The destination of a movement is randomly chosen from four directions. Consumers and resources move a unit length with probabilities $v_{C_i} \Delta t$ and $v_R \Delta t$.

(3) *Forming pairs*: Consumers and resource individuals form chasing pairs within a distance of $r^{(C)}$, while consumers individuals form interference pairs within a distance of $r^{(I)}$.

(4) *Dissociation*: The system undergoes updates at small time step $\Delta t$, ensuring that $d_i \Delta t, k_i \Delta t, d'_{ij} \Delta t \ll 1$ $(i,j=1,2)$. During this process, a random number $\varsigma$ is drawn from a uniform distribution $\mathcal{U}(0,1)$ (where 0 is the low bound and 1 is the upper bound). If $\varsigma < d_i \Delta t$ or $\varsigma < d'_{ij} \Delta t$, the chasing pair or interference pair undergo dissociation into two individuals. One remains in the original position, while the other individual moves just outside the encounter radius at a uniformly distributed random angle[6]. In the case of a chasing pair, when $d_i \Delta t < \varsigma < (d_i + k_i)\Delta t$ is met, the consumer successfully captures the resource, resulting in biomass influx into consumer populations. Conversely, if $\varsigma > (d_i + k_i)\Delta t$ or $\varsigma > d'_{ij} \Delta t$, the chasing pair or interference pair maintains their current status.

(5) *Birth and death*: For each species, we employ two distinct counters with decimal precision to monitor the contributions of both birth and death processes, accumulating in each time step. The increments in the integer part of the counter trigger updates. Specifically, a new member enters the system through the initialization process in a birth action, while an unfortunate individual is randomly selected from the living population in a mortality event.

## 5.2 Simulation details of the main text figures

In Figure 2a, b, e, f: $a_i = 0.05$, $a'_i = 0.0625$, $d_i = 2$, $k_i = 0.22$, $w_i = 0.32$ $(i=1,2)$, $D_1 = 0.055$, $D_2 = 0.058$, $R_0 = 0.15$, $K_0 = 10000$. In Figure 2a, e: $d'_i = 0.2$. In Figure 2b, f: $d'_i = 0.8$. In Figure 2c-d: $a_i = 0.1$, $d_i = 0.1$, $k_i = 0.1$, $w_i = 0.1$ $(i=1,2)$, $D_2 = 0.001$, $\Delta = (D_1 - D_2)/D_2$, $R_0 = 0.05$, $K_0 = 100$. In Figure 2d: $\Delta = 1$. The numerical results in Figure 2a-f were calculated or simulated from Eqs. 1-2.

In Figure 3a-c: $L = 120$, $r^{(C)} = 5$, $r^{(I)} = 5$, $v_{C_i} = 1$, $v_R = 1$, $a_i = 0.0393$, $a'_i = 0.0393$, $d_i = 0.3$, $d'_i = 0.3$, $k_i = 0.1$, $w_i = 0.3$ $(i=1,2)$, $D_1 = 0.0155$, $D_2 = 0.015$, $R_0 = 0.5$, $K_0 = 200$. In Figure 3d-e: $L = 120$, $r^{(C)} = 5$, $r^{(I)} = 5$, $v_{C_i} = 1$, $v_R = 1$, $a_i = 0.3928$, $a'_i = 0.3928$, $d_i = 0.4$, $d'_i = 0.5$, $k_i = 0.1$, $w_i = 0.2$ $(i=1,2)$, $D_1 = 0.0189$, $D_2 = 0.0188$, $R_0 = 0.8$, $K_0 = 400$. The numerical results in Figure 3a-e were simulated from Eqs. 1-2.

In Figure 4a-d: $a_i = 0.01$, $a'_i = 0.0125$, $d_i = 0.8$, $k_i = 0.18$, $w_i = 0.15$, $D_i = 0.015 \times \mathcal{N}(1, 0.16)$ $(i=1,\cdots,20)$; $R_0 = 0.16$, $K_0 = 200000$. The mortality rate $D_i$ $(i=1,\cdots,20)$ is the sole parameter varying with the consumer species, randomly sampled from a Gaussian distribution $\mathcal{N}(\mu, \sigma^2)$ where $\mu$ and $\sigma$ represent the mean and standard deviation, respectively. For ease of notation, we simplify $\mathcal{N}(\mu, \sigma^2)$ as $\mathcal{N}(\mu, \sigma)$ wherever applicable in this paper. The numerical results in Figure 4a-d were simulated from Eqs. 1-2.

Model settings in Figure 5a-b (bird): $a_i = 0.1$, $a'_i = 0.125$, $d_i = 0.5$, $d'_i = 0.6$, $k_i = 0.11$, $w_i = 0.21$, $D_i = 0.015 \times \mathcal{N}(1, 0.3)$ $(i=1,\cdots,20)$; $R_0 = 0.5$, $K_0 = 10^4$. In Figure 5a-b (bat): $a_i = 0.1$, $a'_i = 0.125$, $d_i = 0.5$, $d'_i = 0.5$, $k_i = 0.1$, $w_i = 0.2$, $D_i = 0.013 \times \mathcal{N}(1, 0.34)$ $(i=1,\cdots,40)$; $R_0 = 0.5$, $K_0 = 10^4$. Model settings in Figure 5a-b (fish): $a_i = 0.1$, $a'_i = 0.14$, $d_i = 0.5$, $d'_i = 0.5$, $k_i = 0.1$, $w_i = 0.2$,

$D_i = 0.015 \times \mathcal{N}(1, 0.33)$ $(i = 1, \cdots, 50)$; $R_0 = 0.5$, $K_0 = 10^4$. Model settings in Figure 5a-b (lizard): $a_i = 0.1$, $a_i' = 0.125$, $d_i = 0.5$, $d_i' = 0.6$, $k_i = 0.12$, $w_i = 0.18$, $D_i = 0.012 \times \mathcal{N}(1, 0.36)$ $(i = 1, \cdots, 55)$; $R_0 = 0.6$, $K_0 = 10^4$. In Figure 5a-b, $D_i$ $(i = 1, \cdots, S_C)$ is the only parameter that varies with the consumer species, randomly sampled from a Gaussian distribution $\mathcal{N}(\mu, \sigma)$. The coefficient of variation (CV) for mortality rates (i.e., $\sigma/\mu$) was chosen to be approximately 0.3, specifically determined as the best fit within the range of 0.15-0.43. This interval was derived from experimental findings[7] using the two-sigma rule. The numerical results in Fig. 5 were simulated from Eqs. 1-2. The Shannon entropies of experimental data and simulation results for each community are: $H^{\text{bird}}_{\text{Exp(ODEs,SSA)}} = 2.98(2.67, 2.66)$, $H^{\text{bat}}_{\text{Exp(ODEs,SSA)}} = 3.00(3.04, 2.98)$, $H^{\text{fish}}_{\text{Exp(ODEs,SSA)}} = 3.78(3.42, 3.46)$, $H^{\text{lizard}}_{\text{Exp(ODEs,SSA)}} = 4.05(3.22, 3.08)$. Here, the Shannon entropy is calculated using the formula $H = -\sum_{i=1}^{S_C} P_i \log(P_i)$, where $P_i$ is the probability that a consumer individual belongs to species $C_i$.

**Supplementary References**

**Supplemental Figures**

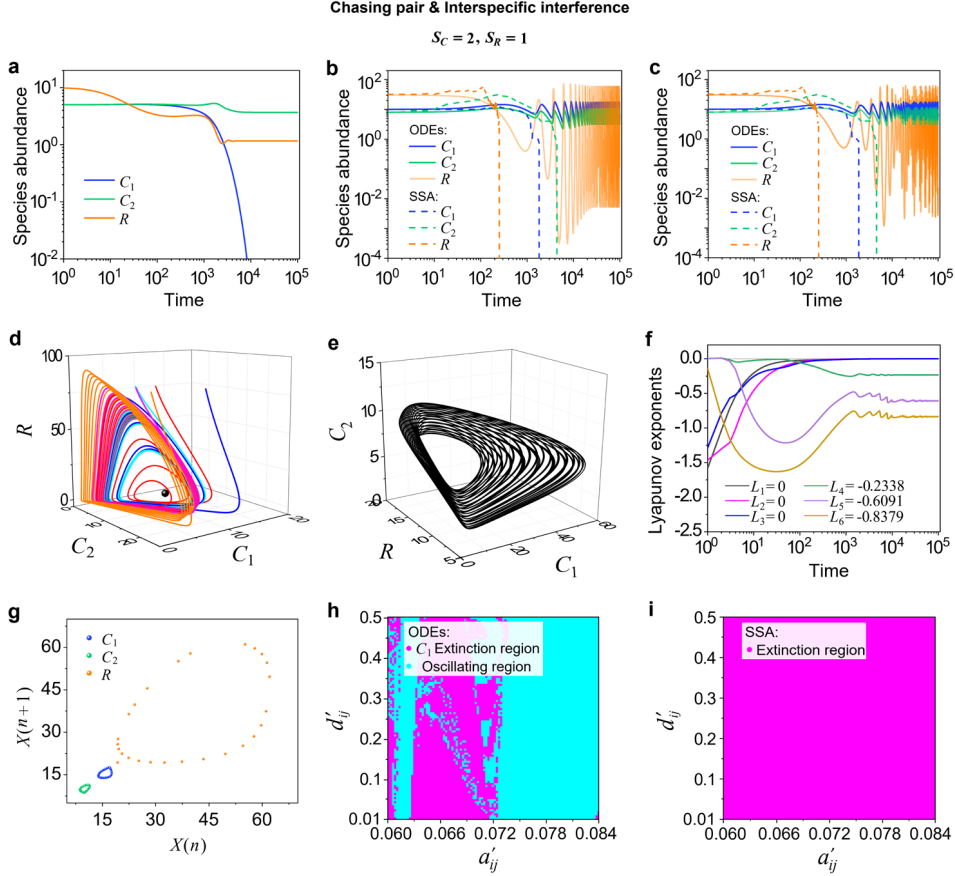

Fig. S1 | Numerical results in the scenario involving chasing pairs and interspecific interference, with $S_C = 2$ and $S_R = 1$. (a-c) Time courses of species abundances simulated with ODEs or SSA[8]. Consumer species cannot coexist at constant population densities, yet they may coexist with time series dynamics in the ODE simulations. However, all coexistence states are vulnerable to stochasticity (see SSA results). (d-e) In a 3D phase space, the ODEs results in (b) and (c) correspond to a limit cycle and a quasi-periodic 3-D torus, respectively. (f-g) The Lyapunov exponents and Poincare map further identify that the dynamics in (c) is a quasi-periodic oscillation. (f) In the Lyapunov exponents spectrum, three exponents are zeros while the rest are all negative, demonstrating a 3-D torus[2]. Here $L_1, L_2, \cdots, L_6$ represent the Lyapunov exponents. (g) The loops in the Poincare map indicate a quasi-periodic oscillation[2]. (h) In the ODEs studies, the cyan region represents oscillating coexistence while the magenta region represents $C_1$ extinction. (i) In the SSA studies, there is no parameter region for species coexistence. The numerical results in (a-i) were simulated from Eqs. 1, 3. In (a): $a_i = 0.05$, $d_i = 0.1$, $k_i = 0.1$, $w_i = 0.05$ $(i=1,2)$; $D_1 = 0.0009$, $D_2 = 0.0007$, $a'_{12} = 0.06$, $d'_{12} = 0.02$, $R_0 = 0.05$, $K_0 = 10$. In (b-i): $a_i = 0.06$, $d_i = 0.2$, $k_i = 0.1$, $w_i = 0.06$ $(i=1,2)$; $D_1 = 0.0009$, $R_0 = 0.05$, $K_0 = 100$. In (b, d): $D_2 = 0.00075$, $a'_{12} = 0.063$, $d'_{12} = 0.05$. In (c, e, f, g): $D_2 = 0.00071$, $a'_{12} = 0.063$, $d'_{12} = 0.05$. In (h-i): $D_2 = 0.00075$.

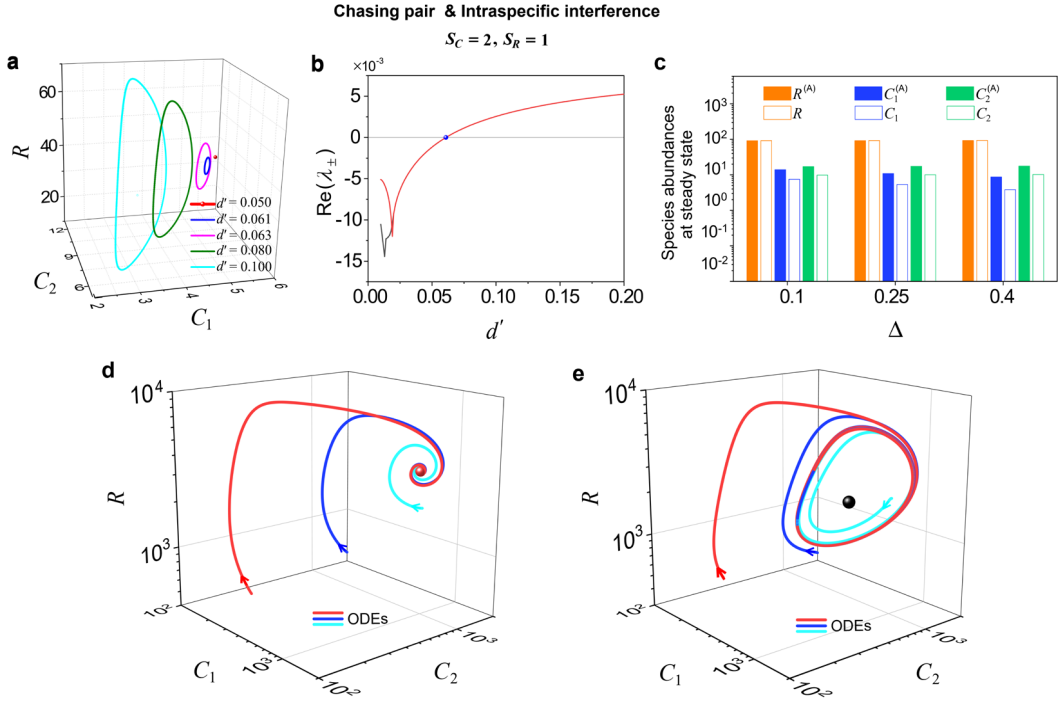

Fig. S2 | Numerical results in the scenario involving chasing pairs and intraspecific interference, with $S_C = 2$ and $S_R = 1$. (a) The phase diagram of a Hopf bifurcation by varying the parameter $d'$ ($d' \equiv d'_1 = d'_2$). (b) The Hopf bifurcation curve. (c) Comparisons between numerical results and analytical solutions of the species abundances at fixed points. Here, $D_i$ ($i = 1, 2$) is the only parameter that varies with the consumer species, and $\Delta = (D_1 - D_2)/D_2$ measures the competitive difference between the two species. Color bars are analytical solutions while hollow bars are numerical results. The numerical results were calculated from Eq. S1.1, while the analytical solutions (marked with superscript "(A)") were calculated from Eqs. S1.4-S1.5. (d) The coexistence state is stable and globally attractive (see Fig. 2a for time courses, ODEs results). (e) The limit cycle is stable. All species coexist in an oscillating way (see Fig. 2b for time courses, ODEs results). The numerical results in (a-e) were calculated or simulated from Eqs. 1-2. In (a-b): $a_i = 0.1$, $a'_i = 0.125$, $d_i = 0.1$, $k_i = 0.1$, $w_i = 0.1$ ($i = 1, 2$); $D_1 = 0.0085$, $D_2 = 0.008$, $R_0 = 0.05$, $K_0 = 100$. In (c): $a_i = 0.05$, $a'_i = 0.06$, $d_i = 0.8$, $d'_i = 0.02$, $k_i = 0.1$, $w_i = 0.1$ ($i = 1, 2$); $D_2 = 0.008$, $R_0 = 0.1$, $K_0 = 100$.

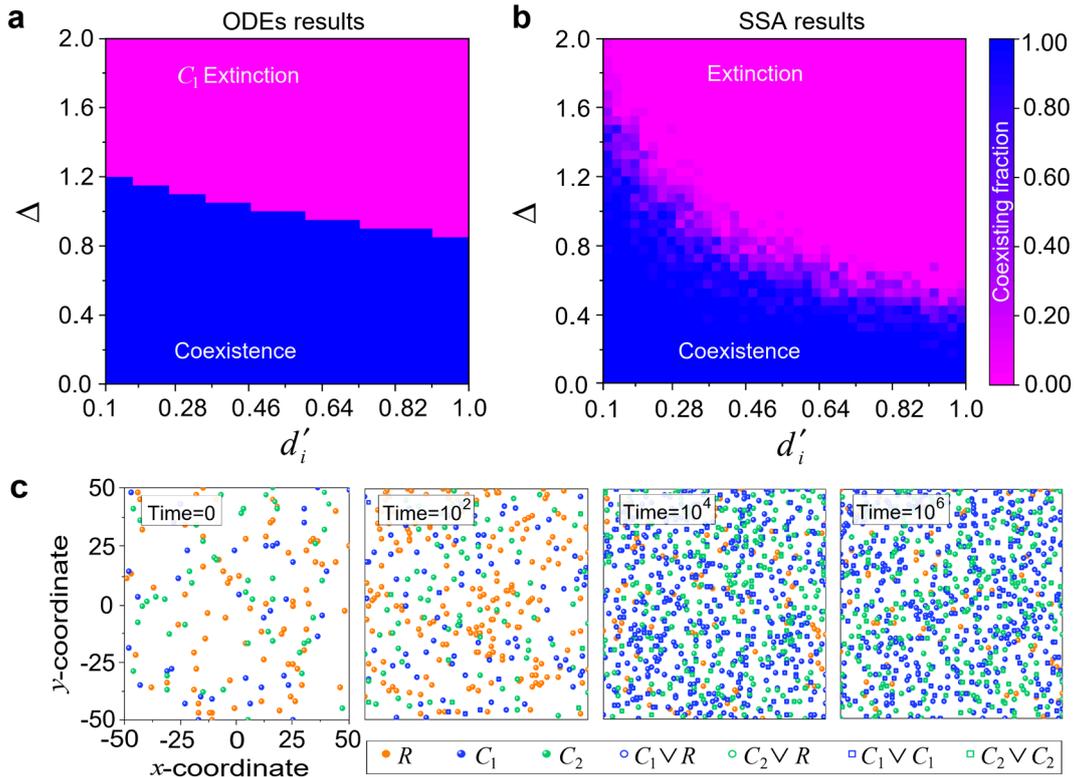

Fig. S3 | Numerical results in the scenario involving chasing pairs and intraspecific interference, with $S_C = 2$ and $S_R = 1$. (a-b) Phase diagrams in the ODEs or SSA studies. Here, $D_i$ ($i=1,2$) is the only parameter varying with the consumer species (with $D_1 > D_2$), and $\Delta = (D_1 - D_2)/D_2$ represents the competitive difference between the two species. (a) and (b) share the same parameter region. The species' coexisting fraction of each pixel in (b) was calculated from 15 random repeats. The numerical results in (a-b) were simulated from Eqs. 1-2. (c) Snapshots of the individual-based modeling (IBM) (see Fig. 3d for time courses). In (a-b): $a_i = 0.05$, $a'_i = 0.625$, $d_i = 0.2$, $k_i = 0.2$, $w_i = 0.32$ ($i=1,2$); $D_2 = 0.025$, $R_0 = 0.15$, $K_0 = 200$. The simulation settings for (c) are identical to those of Fig. 3d-e.

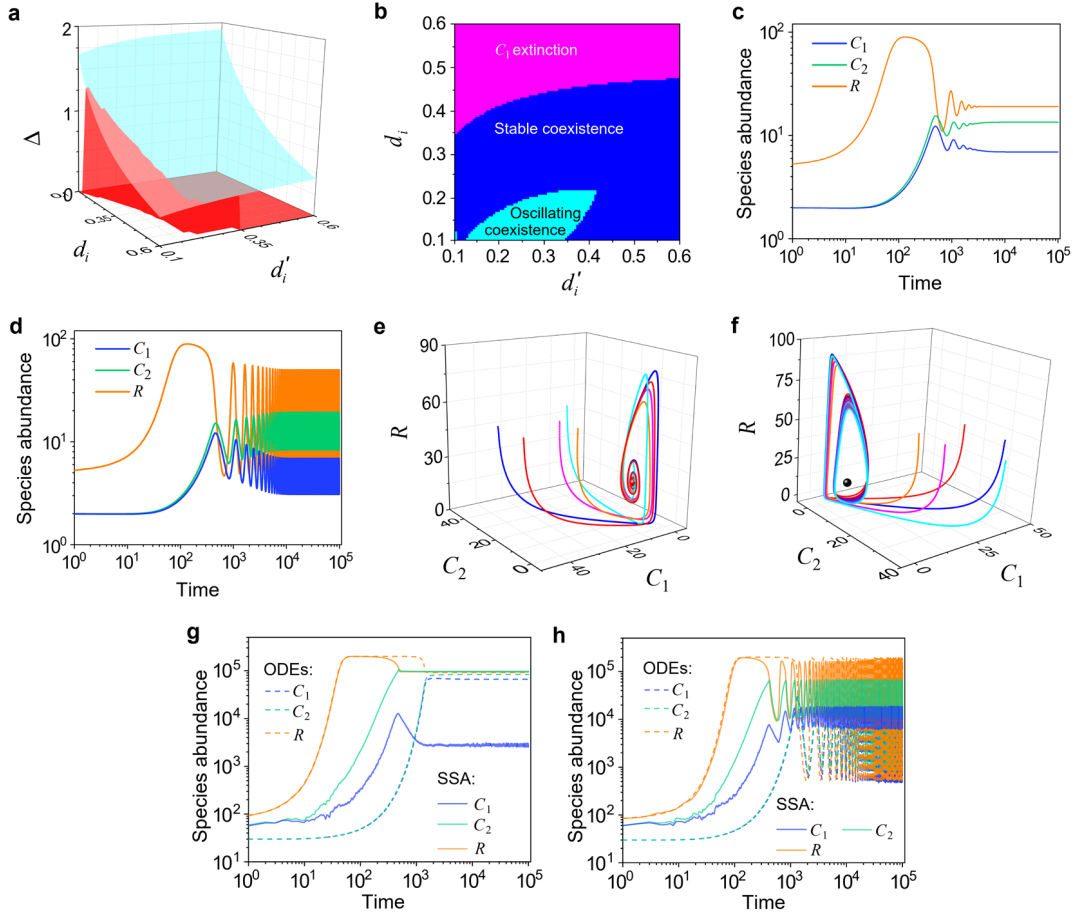

Fig. S4 | Numerical results in the scenario involving chasing pairs and both interspecific and intraspecific interference, with $S_C = 2$ and $S_R = 1$. (a) 3D Phase diagram. Here, $D_i$ ($i = 1, 2$) is the only parameter that varies with the consumer species (with $D_1 > D_2$), and $\Delta = (D_1 - D_2)/D_2$ measures the competitive difference between the two species. The parameter region below the blue surface yet above the red surface represents stable coexistence, while that below the red surface and above $\Delta = 0$ represents unstable fixed points. (b) A transection corresponding to the plane $\Delta = 0.25$ in (a). (c-d) Time courses of the species abundances. Consumer species may coexist at steady state or in an oscillating way. (e) The fixed points (shown in red) are stable and globally attractive (see Fig. S4c for time courses). (f) The fixed point is unstable, and trajectories end in a limit cycle (see Fig. S4d for time courses). (g-h) Time courses of the species abundances with SSA. Consumer species may coexist at steady state or in an oscillating way regardless of stochasticity. The numerical results in (a-h) were simulated from Eqs. 1-3. In (a-f): $a_i = 0.15$, $a_i' = 0.165$, $k_i = 0.11$, $w_i = 0.1$ ($i = 1, 2$); $D_2 = 0.0055$, $a_{12}' = 0.165$, $d_{12}' = 0.6$, $R_0 = 0.075$, $K_0 = 100$. In (b): $D_1 = 0.00688$. In (c, e): $d_i = 0.3$, $d_i' = 0.4$, $D_1 = 0.006$. In (d, f): $d_i = 0.2$, $d_i' = 0.4$, $D_1 = 0.006$. In (g-h): $a_i = 0.05$, $a_i' = 0.065$, $k_i = 0.11$, $d_i = 0.2$, $d_i' = 0.2$, $w_i = 0.12$ ($i = 1, 2$); $D_1 = 0.008$, $D_2 = 0.0079$, $a_{12}' = 0.065$, $d_{12}' = 0.15$, $K_0 = 200000$. In (g): $R_0 = 0.2$. In (h): $R_0 = 0.1$.

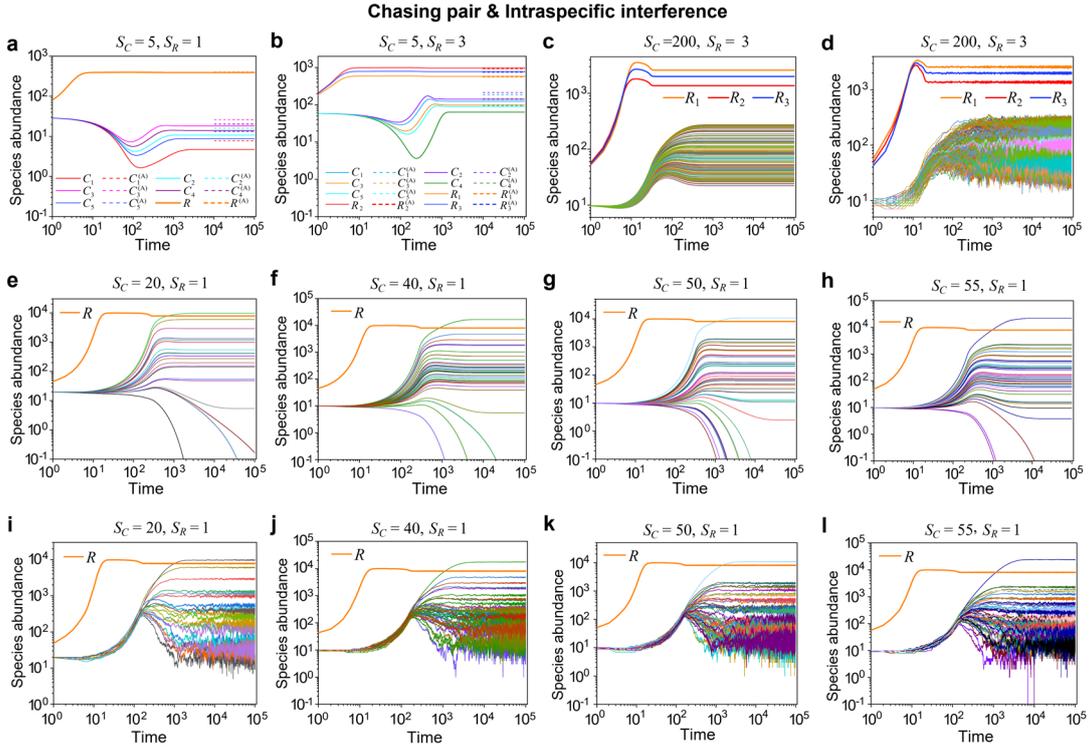

Fig. S5 | Intraspecific predator interference enables a wide range of consumer species to coexist with a handful of resource species. For each model community, $D_i$ $(i=1,\cdots,S_C)$ is the only parameter that varies with the consumer species, which was randomly drawn from a uniform distribution $\mathcal{U}$ or a Gaussian distribution $\mathcal{N}$. (a-l) Time courses of the species abundances simulated with ODEs or SSA. The dotted lines in (a-b) represent the analytical steady-state solutions (marked with superscript "(A)"), which were calculated from Eqs. S2.7a-S2.7e. The time series in (e, i), (f, j), (g, k), and (h, l) correspond to those shown in Fig. 5 (bird, bat, fish, lizard), respectively. The numerical results in (a-l) were calculated/simulated from Eqs. 1-2. In (a): $a_i = 0.04$, $a_i' = 0.056$, $d_i = 0.6$, $d_i' = 0.04$, $k_i = 0.15$, $w_i = 0.45$ $(i=1,2)$; $D_1 = 0.062$, $D_2 = 0.059$, $D_3 = 0.057$, $D_4 = 0.058$, $D_5 = 0.060$, $R_0 = 0.9$, $K_0 = 400$. In (b): $a_{il} = 0.05$, $a_i' = 0.07$, $d_{il} = 1.05$, $d_i' = 0.018$, $k_{il} = 0.016$, $w_i = 0.45$ $(i=1,\cdots,5, l=1,2,3)$; $D_1 = 0.062$, $D_2 = 0.0615$, $D_3 = 0.0639$, $D_4 = 0.066$, $D_5 = 0.0644$, $R_0^{(1)} = R_0^{(2)} = 0.9$, $R_0^{(3)} = 0.95$, $K_0^{(1)} = 600$, $K_0^{(2)} = 1000$, $K_0^{(3)} = 800$. In (c-d): $a_{il} = 0.2$, $a_i' = 0.25$, $d_{il} = 0.4$, $d_i' = 0.2$, $k_{il} = 0.3$, $w_i = 0.3$, $D_i = 0.02 + 0.03 \times \mathcal{U}(0,1)$ $(i=1,\cdots,200, l=1,2,3)$; $R_0^{(1)} = 0.85$, $R_0^{(2)} = 0.95$, $R_0^{(3)} = 0.9$, $K_0^{(1)} = 4000$, $K_0^{(2)} = 2000$, $K_0^{(3)} = 3000$. The simulation settings for (e-l) are identical to those of Fig. 5.